\documentclass[double-spaced]{llncs}

\usepackage{amssymb}
\usepackage{xspace}
\usepackage[english]{babel}
\usepackage{amsmath}
\usepackage{amsopn}
\usepackage{latexsym}
\usepackage{textcomp}
\usepackage{epsfig}
\usepackage{subfigure}
\usepackage{enumerate}
\usepackage{tabularx}
\usepackage[plainpages=false]{hyperref}
\usepackage[figure,table]{hypcap}
\usepackage{multirow}
\usepackage{todonotes}
\usepackage{color, colortbl}
\usepackage{paralist}
\usepackage{soul}
\usepackage{pdflscape}

\usepackage{amsmath}
\usepackage{algorithm}
\usepackage[noend]{algpseudocode}
\makeatletter
\def\BState{\State\hskip-\ALG@thistlm}
\makeatother

\newcommand{\remove}[1]{}
\newcommand{\sys}{\textsf{GiViP}\xspace}

\title{\sys: A Visual Profiler for Distributed Graph Processing Systems
\thanks{We thank Maria Elisa Ganci for her contribution in the development of  \sys.}
}
\author{
Alessio Arleo,
Walter Didimo,
Giuseppe Liotta,
Fabrizio Montecchiani,
}

\date{}

\institute{
Universit\`a degli Studi di Perugia, Italy
\\
\email{alessio.arleo@studenti.unipg.it} \\
\email{\{walter.didimo,giuseppe.liotta,fabrizio.montecchiani\}@unipg.it}
}

\begin{document}
\maketitle

\begin{abstract}
Analyzing large-scale graphs provides valuable insights in different application scenarios.
%
While many graph processing systems working on top of distributed infrastructures have been proposed to deal with big graphs, the tasks of profiling and debugging their massive computations remain time consuming and error-prone. This paper presents \sys, a visual profiler for distributed graph processing systems based on a Pregel-like computation model. \sys captures the huge amount of messages exchanged throughout a computation and provides an interactive user interface for the visual analysis of the collected data. We 
show how to take advantage of \sys to detect anomalies related to the computation and to the infrastructure, such as slow computing units and anomalous message patterns. 
\end{abstract}

\section{Introduction}\label{se:introduction}
The analysis of large-scale graphs provides valuable insights in different application scenarios, including social networking, crime detection, content ranking and recommendations (see, e.g,~\cite{DBLP:journals/pvldb/ChingEKLM15,facebook,DBLP:conf/sigmod/MalewiczABDHLC10,paypal}). On the other hand, graph computations are often difficult to scale and parallelize, due to the inherent interdependencies within graph data. Furthermore, graph algorithms are usually iterative and hence poorly suited for popular Big Data processing systems such as Hadoop/MapReduce (see, e.g.,~\cite{DBLP:journals/cse/Cohen09,DBLP:journals/ppl/LumsdaineGHB07}). In response to these shortcomings, new frameworks based on the Think-Like-A-Vertex (TLAV) programming model have been proposed, such as Google's Pregel~\cite{DBLP:conf/sigmod/MalewiczABDHLC10} and its open source counterpart Apache Giraph~\cite{DBLP:journals/pvldb/ChingEKLM15}. The idea behind the TLAV model is to provide a common vertex-centric programming interface, abstracting from low-level details of the distributed infrastructure. Graph processing systems based on the TLAV model outperform general purpose Big Data processing systems by improving locality and by demonstrating linear scalability~\cite{DBLP:journals/csur/McCuneWM15}. In view of their effectiveness, these systems are being adopted by a growing number of applications. For example Apache Giraph is used in the contexts of social networking~\cite{facebook}, fraud detection~\cite{paypal}, and network visualization~\cite{DBLP:conf/gd/ArleoDLM16,DBLP:journals/isci/ArleoDLM17}.

While many graph processing systems working on top of modern distributed infrastructures have been proposed to deal with large graphs, the tasks of profiling and debugging their massive computations remain time consuming and error-prone~\cite{DBLP:conf/icse/GulzarIYTCMK16,DBLP:conf/sigmod/SalihogluSKTW15}. Low-level profiling systems for distributed architectures exist~\cite{hprofiler,statsd}, but none of them is tailored to the needs of TLAV frameworks (or other types of distributed graph processing systems). For example, Hadoop Profiler~\cite{hprofiler} is designed to analyze CPU workloads of Apache Hadoop clusters~\cite{hadoop}, but it disregards the interaction between pairs of computing units, which is crucial in a TLAV framework. Indeed, algorithms written for TLAV-based graph processing systems usually rely on slim user-defined functions that do not require much CPU resources, but they may require huge numbers of messages and/or iterations to propagate the results of a local computation throughout the graph. A classical example is the TLAV implementation of the PageRank algorithm, which requires each vertex to iteratively execute a simple computation and to communicate the output to all its neighbors, until convergence is achieved~\cite{DBLP:conf/sigmod/MalewiczABDHLC10}. Moreover, anomalies related to the distributed infrastructure may yield to unbalanced partitions of the input graph over the computing units which, in turn, leads to overloaded links in the distributed infrastructure. Similarly, a buggy implementation of an algorithm may yield to anomalous message patterns. 

\begin{figure}[t]
\centering
\includegraphics[width=\columnwidth]{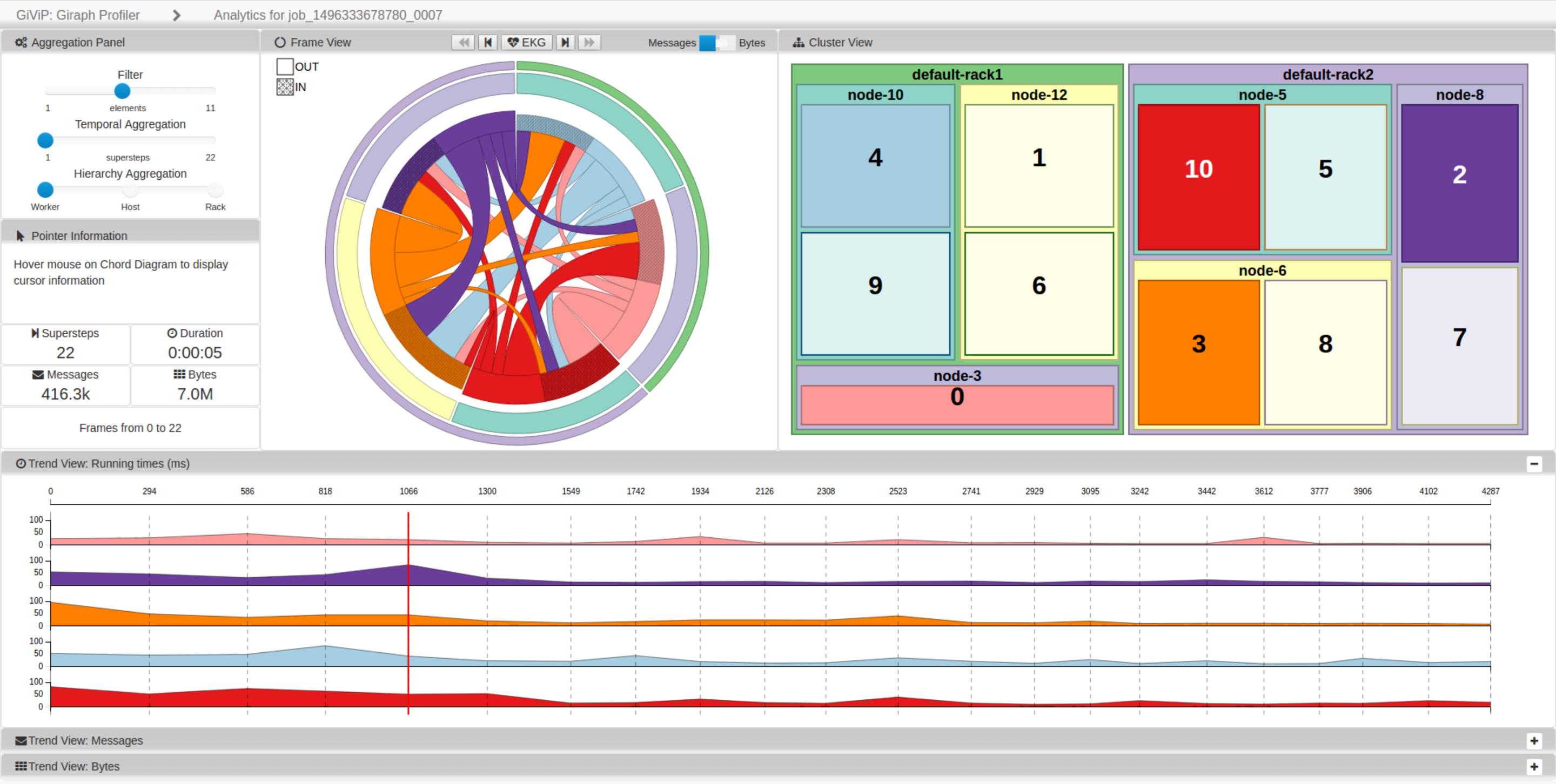}
\caption{The graphical interface of \sys.\label{fig:givip}}
\end{figure}

\noindent{\bf Contribution.} In this paper we tackle the challenge of profiling massive computations that run on top of a TLAV-based graph processing system, and we provide a publicly available implementation of our approach\footnote{http://givip.graphdrawing.cloud/}, called \sys, for Apache Giraph. Figure~\ref{fig:givip} shows a screenshot of the graphical interface of \sys. The system collects the networked data related to messages exchanged by pairs of computing units throughout a specific computation, constructs suitable aggregations of these data, and presents to the user an interactive visual interface for exploring them. 
To demonstrate the effectiveness of our approach, we discuss key usage scenarios of \sys in terms of resource profiling and detection of both computation- and infrastructure-related issues, such as overloaded computing units and anomalous message patterns.
For reasons of space some material has been omitted and can be found in the appendix.

%

\section{Background and Related Work}\label{se:background-related}

\noindent{\bf Background.} The \emph{Think-Like-A-Vertex (TLAV) programming model} provides a common vertex-centric programming interface, abstracting from low-level details of the  computation and of the distributed infrastructure.  
Assuming (with no restrictions) that the input graph is directed, a \emph{user-defined function} aims at updating the internal value of the vertex and/or of its outgoing edges. It takes as input data from the incoming edges of the vertex, while its output is communicated through the outgoing edges. Thus, each vertex exchanges \emph{messages} only with its neighbors. Google's Pregel~\cite{DBLP:conf/sigmod/MalewiczABDHLC10} was the first published implementation of a TLAV framework. It is based on the Bulk-Synchronous Programming (BSP) model~\cite{DBLP:journals/cacm/Valiant90}, which splits the computation into iterations called \emph{supersteps}, with synchronization barriers occurring between consecutive supersteps. At each superstep the user-defined function is executed over the vertices of the graph, and the messages sent by a vertex during a superstep are received by its neighbors at the beginning of the next superstep. 
The computation halts after a number of rounds, or when a halting condition is met.
%
Apache Giraph~\cite{DBLP:journals/pvldb/ChingEKLM15} is a popular Java-based TLAV framework built on  Apache Hadoop~\cite{hadoop} and originated as the open source counterpart of Pregel. Giraph exhibits additional features with respect to Pregel, but it is still based on the BSP model. A fundamental ingredient of large-scale graph processing systems is a preliminary partitioning operation that splits the input graph into parts  assigned to  different computing units. Good partitions often lead to improved performance, but expensive partitioning strategies may end up dominating the processing time. In Giraph, a basic computing unit is called \emph{worker}, and each computer, or \emph{host}, can run multiple  workers. In large clusters, hosts are grouped into \emph{racks}. Giraph provides a default hash-based partitioning algorithm to assign each vertex of the input graph to a worker. Different strategies can be employed by overriding suitable methods of the library. 
We point to the survey by McCune et al.~\cite{DBLP:journals/csur/McCuneWM15} for further references and explanations about TLAV frameworks. In particular, Apache Hama~\cite{DBLP:conf/cloudcom/SeoYKJKM10} and GPS~\cite{DBLP:conf/ssdbm/SalihogluW13} are Pregel-like systems, hence our approach can be adapted for them.


\smallskip\noindent{\bf Debuggers, profilers, and monitoring tools.} 
While modern distributed platforms transparently handle the hassles related to the distributed infrastructure, debugging and profiling computations, as well as monitoring and optimizing the underneath infrastructure, remain challenging tasks. Hadoop Profiler~\cite{hprofiler} is a tool to analyze CPU workloads for Apache Hadoop clusters. The {statsd-jvm-profiler}~\cite{statsd} enables the analysis of memory usage, garbage collection, and the aggregate execution time of each function within Apache Hadoop clusters. Both these tools work at low-level, without distinguishing between concurrent computations running on the same cluster. 
We also mention high performance computing (HPC) profilers such as Gprof~\cite{graham2004gprof} and VTune~\cite{reinders2005vtune}, which sample the execution of a computation and analyze the time spent on each part of the code.
BigDebug~\cite{DBLP:conf/icse/GulzarIYTCMK16} is a tool offering interactive, real-time debugging primitives for computations running on Apache Spark~\cite{spark,Zaharia:2012:RDD:2228298.2228301}, an in-memory engine for Apache Hadoop. Graft~\cite{DBLP:conf/sigmod/SalihogluSKTW15} offers a graphical interface to debug TLAV programs, and it is implemented for Apache Giraph. None of these tools offers resource profiling features. CloudGazer~\cite{DBLP:conf/apvis/StitzGKS15} is a visualization system that allows users to monitor cloud-based networks. This system has provided valuable inspiration for our work but its focus is different from ours, as it is directed towards the optimization of cloud-based infrastructures in order to reduce energy consumption and to increase the quality of service. 

\smallskip\noindent{\bf Time series visualizations.} Profiling a computation involves the analysis of time-varying parameters. Classic charts for time-series data include line charts~\cite{playfair}, small multiples~\cite{tufte1983visual}, stacked graphs~\cite{Byron}, horizon graphs~\cite{Saito}, and braided graphs~\cite{DBLP:journals/tvcg/JavedME10} (see also~\cite{HeerACM}). Javed et al.~\cite{DBLP:journals/tvcg/JavedME10} compared these types of visualizations in a user study with local and global tasks on samples with up to $8$ simultaneous time series. They observed that shared-space visualizations excel at comparisons with a local visual span, while split-space techniques are more robust against high numbers of concurrent time series for tasks that need large visual spans. More compact iconic representations can also be used when dealing with many simultaneous time series, at the expenses of  a less intuitive temporal encoding; see, e.g., the survey by Ward~\cite{Ward2008} and the user study by Fuchs et al.~\cite{DBLP:conf/chi/FuchsFMBI13}. Also, several application-driven systems have been proposed that make use of ad-hoc visualizations. Examples are: ThermalPlot~\cite{thermalplot}, for the visualization of multi-attribute time-series data highlighting significant developments over time; CloudLines~\cite{cloudlines2011}, for time-based representations of large and dynamic event data sets;  LiveRAC~\cite{liverac2008}, for the visualization of large collections of system management time-series data with hundreds of parameters;  ThemeRiver~\cite{themeriver}, for visualizing thematic variations over time within a large collection of documents; LifeLines~\cite{Plaisant1996}, for representing personal histories.

\smallskip\noindent{\bf Dynamic graph drawing.} In \sys the communication among workers is conveniently modeled as a graph whose edges are weighted based on the amount or the size of messages exchanged between pairs of workers during a superstep. Hence, our problem intersects the rich literature on dynamic graph drawing (see, e.g.,~\cite{DBLP:journals/cgf/BeckBDW17,DBLP:journals/tvcg/BurchVBDW11,DBLP:journals/jgaa/CrnovrsaninCM17,DBLP:journals/tvcg/FrishmanT08}). Nonetheless, the topology of our communication graph is unlikely to change over time, as each worker communicates with workers that manage the neighbors of its vertices, regardless of the superstep.

\section{The \sys System}\label{se:system}
%

\subsection{Tasks and requirements}\label{sse:tasks}

The tasks that guided the design of \sys are conceived having in mind the analysis of the resources used by computations running on top of Pregel-like graph processing systems; thus, they substantially differ from the common objectives of low-level distributed profilers. The main tasks are as follows.

\begin{inparaenum}[\bf{T}1]

\item \emph{Analyze the performance trend of a computation in terms of running time and traffic load.} This task is relevant to evaluate the scalability of a distributed algorithm and to detect possible bottlenecks. High running times may be alleviated by scaling up the resources of the cluster; at the same time, adding computational units may even increase the traffic load (as it increases the input fragmentation). Also, peaks of resources may be caused by software or hardware faults, and a deeper inspection of the data may shed more light on the problem.\label{t:performance}

\item \emph{Analyze the traffic between pairs of computing units (workers, hosts, racks).} This is useful to detect overloaded links at different levels of the cluster hierarchy, and to estimate the quality of the graph partitioning algorithm. Note that links between racks are usually slower than links between hosts in the same rack, which are in turn slower than links between workers in the same host.\label{t:traffic}


\item \emph{Analyze data aggregated at different computing scale and time scale.} Aggregating data at different computing scales is needed because the size of a cluster can vary from a few hosts in the same rack, up to many hosts within multiple racks. By aggregating data at different time scales we mean the possibility of aggregating sequences of supersteps. This is particularly useful for executions that span hundreds or thousands of supersteps. The number of supersteps taken by an execution usually depends on several variables such as the structural properties of the input graph, the type of algorithm, and the halting condition.
\label{t:aggregate}

\end{inparaenum}

\smallskip

\noindent We also considered two requirements aimed at simplifying the usage of the system:
\begin{inparaenum}[\bf{R}1]
\item \emph{Avoid user code instrumentation.} While distributed debuggers  often require specific instructions to be incorporated in the user code (see, e.g.,~\cite{DBLP:conf/sigmod/SalihogluSKTW15}), this is commonly avoided in distributed profilers. This feature facilitates the portability of the code in production environments, as the profiler can be switched off without recompiling the user code.  
\label{r:no-code-instr}
\item \emph{Allow remote access to the user interface.} This is essential when the user has no direct access to the computing platform (e.g., when using PaaS products such as Amazon EC2), but instead uses a remote connection or a Web interface to access the cluster.
\label{r:remote-access}
\end{inparaenum}

\subsection{Data model and data aggregation}\label{sse:data}

We now describe how data are organized in \sys and how they can be aggregated to support scalability in the visual interface. 

\smallskip
\noindent{\bf Data model.} The inclusion relationships between workers, hosts, and racks (see Section~\ref{se:background-related}) are represented by an inclusion tree $T$, which does not change over time. A Giraph computation, called \emph{job}, is spread over a sequence of $k > 0$ synchronized supersteps. For each superstep $i$ (for $i = 1, \dots, k$), starting at instant $s_i$, the data collected by \sys are modeled as a weighted directed graph (digraph) $G_i=(V_i,E_i)$. Each vertex $v$ of $G_i$ represents a worker and has a weight $t_i(v)$, denoting the time taken by the worker to complete its task in superstep $i$. The synchronization barriers between supersteps imply that $s_i$ + $max_{v \in V_i}\{t_i(v)\} \leq s_{i+1}$.  Also,  each directed edge $(u,v)$ has two weights, $m_i({uv})$ and $b_i({uv})$, denoting the number of messages and their total size (in bytes) sent from $u$ to $v$ during superstep $i$, respectively. 

\smallskip
\noindent{\bf Data aggregation.} \sys allows two types of data aggregation. \emph{Temporal aggregation} consists of grouping consecutive supersteps in a single \emph{frame}. Let $s_i$ and $s_j$ ($i \le j$) be the first and the last superstep of a frame $f_{ij}$. The system computes a weighted digraph $G_{ij}=(V_i \cup V_{i+1} \cup \dots \cup V_j, E_i \cup E_{i+1} \cup \dots \dots \cup E_j)$. For example, if a computation takes $10,000$ supersteps, a temporal aggregation with $100$ supersteps per frame results in a sequence of $100$ digraphs. The weight of each vertex $v$ of $G_{ij}$ is $t_{ij}(v)=\sum_{z=i}^{j}{t_z(v)}$, and for each edge $(u,v)$ of $G_{ij}$, we have $m_{ij}({uv})=\sum_{z=i}^{j}{m_z({uv})}$ and $b_{ij}({uv})=\sum_{z=i}^{j}{b_z({uv})}$. 

\emph{Hierarchy aggregation} merges workers based on their membership in the same host or rack. Aggregating data in a hierarchical fashion is a well established method to alleviate visual clutter and to support scalability~\cite{Elmqvist:2010}. Consider a weighted digraph $G_{ij}$ (possibly with $i=j$).  A hierarchy aggregation at the \emph{host level} computes a weighted digraph  $G_{ij}^H$  as follows. For each host $h \in T$ we have a vertex $v$ in $G_{ij}^H$, whose weight $t^H_{ij}(v)$ equals the sum of the weights of all vertices of $G_{ij}$ that belongs to $h$. Similarly, we have an edge $(u,v)$ in $G_{ij}^H$ if there is at least an edge in $G_{ij}$ between a vertex in the host of $u$ and a vertex in the host of $v$. The weights $m^H_{ij}(uv)$ and $b^H_{ij}(uv)$ are computed as the sum of the corresponding weights over all edges between a vertex in the host of $u$ and a vertex in the host of $v$. Analogously, a hierarchy aggregation at the \emph{rack level} computes a graph  $G_{ij}^R$  by aggregating workers in the same rack. A hierarchy aggregation at the \emph{worker level} trivially corresponds to $G_{ij}^W=G_{ij}$.

In what follows, for a weighted digraph $G_{ij}$ we assume that $i \le j$. If $i=j$, then no temporal aggregation has been performed. To simplify the notation,  we may omit the superscript (W, H, or R) that specifies the hierarchy aggregation level, if this is not relevant for the discussion and does not create ambiguities.


\subsection{Visualization paradigm and interface}\label{sse:visualization}

The interface of \sys allows users to interactively explore the networkeddata associated with a  computation. The interface is divided into four main views, which we call \emph{Aggregation Panel}, \emph{Cluster View}, \emph{Trend View}, and \emph{Frame View} (see Fig.~\ref{fig:givip}). 
The Trend View and the Frame View mainly support tasks {\bf T}\ref{t:performance} and {\bf T}\ref{t:traffic}, respectively. The Aggregation Panel supports task {\bf T}\ref{t:aggregate}. The Cluster View conveys the hierarchical structure of the computing cluster and is used to filter elements of this hierarchy. The three views are coordinated and highly interactive.  
Each worker is associated with a unique color, which is consistently used in all views. We used color schemes offered by the D3.js library~\cite{d3}.

\smallskip
\noindent{\bf Aggregation Panel.} It contains controls that have impact on both the Trend View and the Frame View. 
A temporal aggregation can be performed by using a slider to set the size of each frame.  A hierarchy aggregation  can be set by means of a three-state switch. 
In addition, the user can filter the computing units based on the total amount of messages they exchange, so to hide those units that have a smaller impact in terms of traffic load.
Finally, this panel contains some high-level statistics such as the total running time and the number of supersteps taken by the computation, and the total number of exchanged messages and bytes.

\smallskip
\noindent{\bf Cluster View.} Interacting with this view allows focusing only on a subset of computational units, by filtering out workers, hosts, and racks. Filtered workers disappear from both the Frame and the Trend View. If a host (rack) is filtered out, then all its workers (hosts) are filtered out. 
The inclusion tree $T$ is shown by means of a squarified treemap~\cite{DBLP:conf/vissym/BrulsHW00}. By clicking on a tile, the corresponding computational unit is filtered in or out based on its current state. The size of a tile is proportional to the number of vertices of the input graph assigned to the corresponding computational unit. This is helpful in two ways. First, the user can decide to filter those units that contain fewer vertices. Second, the user has an immediate feeling of whether the graph partitioning algorithm produced a balanced partition or not. Recall that Giraph's default partitioning algorithm guarantees balanced partitions, but different strategies can be employed to optimize other criteria, such as minimizing inter-worker links~\cite{DBLP:conf/icdcs/VaqueroCLM14}.

\smallskip
\noindent{\bf Trend View.} 
\begin{figure}[t]
\centering
\includegraphics[width=1\columnwidth]{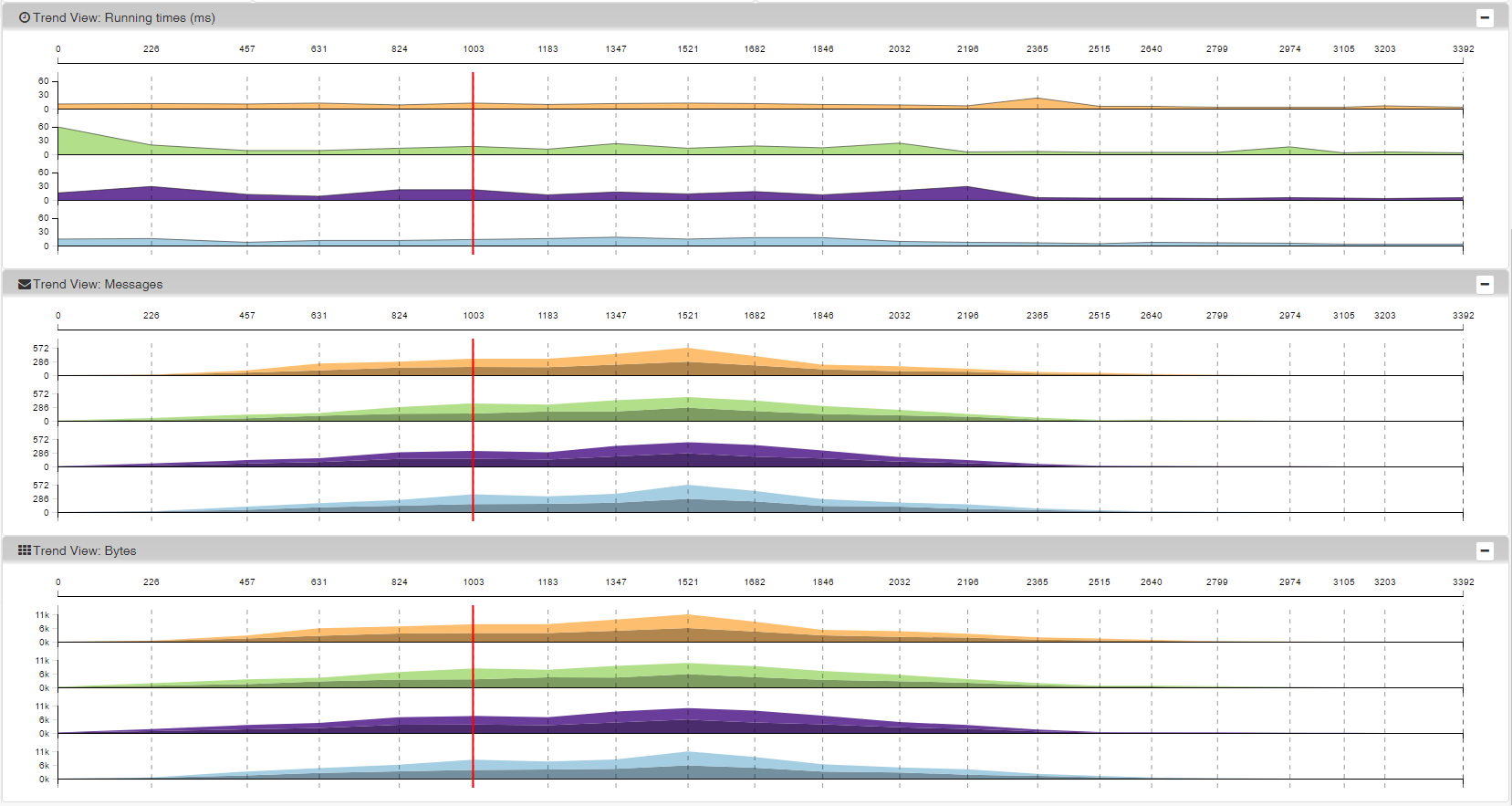}
\caption{The Trend View.}\label{fig:trend-view}
\end{figure}
For each computing unit, this view shows the evolution throughout the computation of running time, number of exchanged messages, and amount of exchanged bytes. (A computing unit is a worker, a host, or a rack, depending on the hierarchy aggregation level.) We encode this information as a set of three small multiples~\cite{tufte1983visual}, vertically stacked and with a shared time axis, see Fig.~\ref{fig:trend-view}. We recall that Javed et al.~\cite{DBLP:journals/tvcg/JavedME10} experimentally observed that split-space visualizations are particularly robust against various concurrent time series for tasks that need large visual spans, which is exactly our setting ({\bf T}\ref{t:performance}). 
The first small multiple shows the running time over all the computation frames. Each single chart is an area chart showing the evolution for the corresponding computing unit. The second small multiple shows the number of messages exchanged over all computation frames. Each single chart is a stacked area chart (also known as stream graph) that shows both the incoming and the outgoing messages of the corresponding computing unit, and thus which also conveys the total number of messages. The incoming messages are depicted with a regular texture to darken the original color assigned to the computing unit. Distinguishing between incoming and outgoing messages is useful because each worker is responsible only for the outgoing edges incident to its vertices, while the incoming edges play a role in the amount of messages that will be received in the next frame. The third small multiples is similar to the previous one but the traffic load is measured in terms of bytes. Each of the three small multiples is enclosed in a collapsible panel. 
Finally, the shared time axis is paginated and initialized responsively with a number of frames per page to guarantee an adequate resolution. As a rule of thumb, a display with $1920 \times 1080$ px allows up to $50$ frames, while $20$ frames guarantees a pleasant distribution of the labels.

\smallskip
\noindent{\bf Frame View.} 
\begin{figure}[t]
\centering
\subfigure[]{\includegraphics[width=0.45\columnwidth]{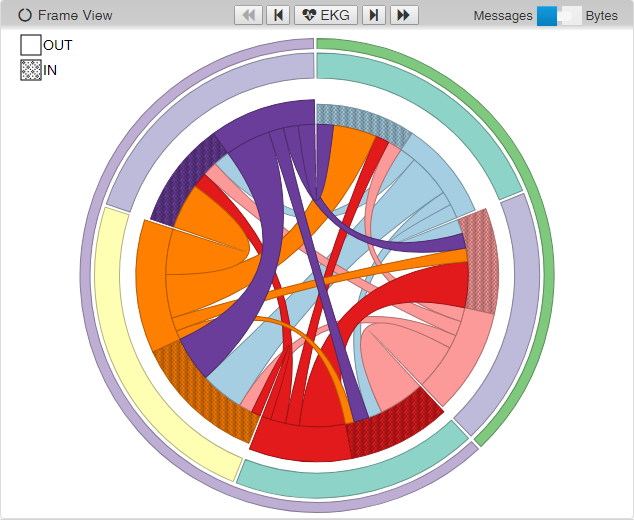}\label{fig:frame-view-worker}}\hfil
\subfigure[]{\includegraphics[width=0.45\columnwidth]{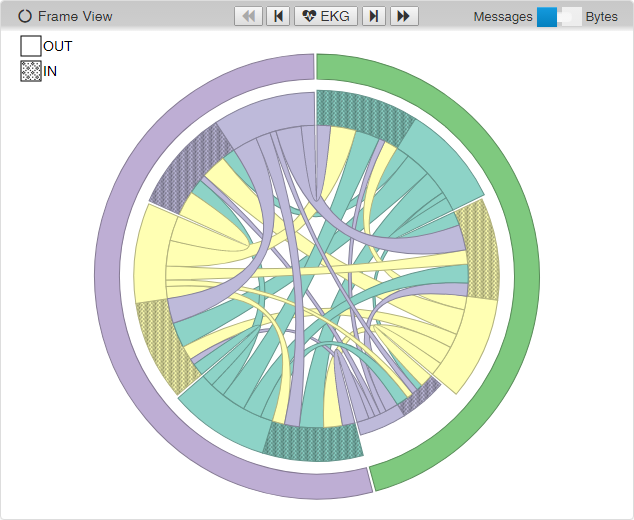}\label{fig:frame-view-host}}
\caption{The Frame View, hierarchy aggregation at (a) worker and (b) host level.}\label{fig:frame-view}
\end{figure}
Here we depict the traffic load between pairs of computing units. Let $G_{ij}$ be the digraph corresponding to a frame $f_{ij}$. As discussed in Section~\ref{se:background-related}, the topology of $G_{ij}$ does not depend on the specific frame $f_{ij}$. Indeed, as we observed in our experiments, $G_{ij}$ is usually a complete graph, especially if the hierarchy aggregation is set to the host or rack level. On the other hand, the edge weights may significantly differ depending on the frame. These observations motivate a network visualization method that privileges the user mental map preservation~\cite{DBLP:journals/tvcg/ArchambaultPP11,DBLP:conf/gd/PurchaseHG06}, and that is conceived to effectively encode edge weights. We implemented an enhanced version of the chord diagram available in~\cite{d3}, as shown in Fig.~\ref{fig:frame-view}. A chord diagram is a circular layout in which the vertices of the graph are arranged as thick circle arcs, and the edges are shown with ribbons connecting pairs of arcs. The size of a ribbon encodes the quantitative information associated with the corresponding edge, and thus each circle arc is long enough to accommodate the ends of its ribbons. Chord diagrams are effectively adopted in various contexts such as comparative genomics~\cite{circos}, urban mobility trajectories~\cite{Gabrielli2014}, and others~\cite{circosweb}. Also, they can be extended to support hierarchical data sets (see, e.g.,~\cite{DBLP:conf/grapp/ArgyriouSV14,DBLP:journals/tvcg/Holten06}), as in our case. We use concentric circles to encode the hierarchy levels. Circle arcs representing workers (hosts) in the same host (rack) appear consecutively around the circle. If the hierarchy aggregation is set at the worker level, then the three levels of the hierarchy are simultaneously shown; see Fig.~\ref{fig:frame-view-worker}. If the data are aggregated at host or rack level, then only two levels or one level are shown, respectively;~see~Fig.~\ref{fig:frame-view-host}. 
The main novelties introduced by our enhanced chord diagram are: $(i)$ the use of heuristics for crossing minimization inspired by the literature on circular layouts (see, e.g.,~\cite{DBLP:conf/wg/BaurB04,DBLP:journals/tcs/DehkordiEHN16,cise,Six1999}), and $(ii)$ a bimodal orientation of the edges in which the incoming and the outgoing edges of each vertex form two contiguous intervals~\cite{DBLP:books/ph/BattistaETT99}.

\smallskip\noindent{\em Edge crossing minization.} Edge crossings are a form of visual clutter that deteriorates the readability of a drawing~\cite{p-eivsg+-00,pca-eeagl+-02,wpcm-cmga-02}. We use a variant of the heuristic by Baur and Brandes~\cite{DBLP:conf/wg/BaurB04} to minimize edge crossings (the optimization problem is \textsc{NP}-complete~\cite{masuda1987np}); it deals with the constraints imposed by the inclusion tree $T$ and with (dynamic) edge weights. Our algorithm takes as input the graph $G_{1k}$, where $k$ is the number of supersteps of the computation, and computes a unique circular order of the vertices, used for the visualization of all graphs $G_{ij}$. This is crucial for the user mental map preservation, especially when the visualization changes due to filtering or aggregations. 

\smallskip\noindent{\em Bimodal orientation.} In the chord diagram, the orientation of an edge is encoded by coloring its ribbon with the same color as the source  vertex. In addition, we split the circle arc of a vertex into two intervals, one for the incoming edges and one for the outgoing. The length of each interval reflects the total weight of the corresponding edges, which facilitates the comparison between incoming and outgoing traffic at a computing unit.
To avoid crossings between adjacent edges, the outgoing edges of a vertex always follow the incoming edges in clockwise order. The interval for the incoming edges is filled with a regular pattern to darken its original color (as in the Trend View). Furthermore, in our chord diagram, a self-loop is encoded by thickening its vertex (circle arc) proportionally to its weight; this helps to understand the amount of traffic within the same unit. 

\smallskip
\noindent{\bf Interaction.} Every aggregation or filtering operation is immediately reflected in all views. Changes in the Trend and Frame Views are smoothed by animated transitions, which help in preserving the user mental map. The time axis of the Trend View is anchored with a slider to browse the frames of the computation. When the user releases the cursor of the slider, the chord diagram smoothly changes the width of its ribbons, so to highlight significant changes. By mouse hovering on the various visualizations, details are immediately shown through pop-ups. For example, by hovering a ribbon of the chord diagram, the number of messages (and bytes) associated with the edge is displayed, or by hovering an area chart, the corresponding value of the diagram is shown.

\subsection{Architecture and implementation notes}\label{sse:architecture}

The architecture of \sys is composed of two main modules. The \textsc{Message Sniffer} collects all data that need to be analyzed. It is realized as a patch for Giraph's source code and can be  deployed without user code instrumentation ({\bf R}\ref{r:no-code-instr}). The data are collected asynchronously 
so to minimize the impact of this module on the computation. Some experiments (on $20$ computations) showed that using our patch does not slow down a Giraph job by more than $36 \%$, and only by $7.5 \%$ on average.
As a comparison, other systems to monitor parallel and distributed algorithms have an overhead around 5\%~\cite{hpctoolkitwebsite,braunddtrace}.
Although \sys is not meant to be used in production environments, these numbers suggest the profiling activity does not seriously affect  the running time of a computation. The \textsc{Visual Analyzer} has a Java back-end that aggregates and stores the collected data in a MySQL database, and that provides a RESTful API to access the data. The front-end runs in a Web browser ({\bf R}\ref{r:remote-access}) and implements the GUI of \sys. It is coded in HTML/CSS/Javascript and exploits the D3.js~\cite{d3} library.

\section{Usage Scenarios}\label{se:evaluation}

We discuss the effectiveness of \sys in key scenarios covering all tasks of Section~\ref{sse:tasks}. We used two clusters, depending on the experiment. One is an Amazon EC2 cluster with $1$ rack, $10$ hosts, and $20$ workers. The other is a cluster of commodity machines at our university with $1$ rack, $6$ hosts, and $11$ workers. 

\smallskip\noindent{\bf Scenario 1: Resource profiling.} 
Distributed algorithms are characterized by the trend of the performance parameters throughout a computation. This trend can be regarded as the ``heartbeat'' of the algorithm, as it is only partially affected by the input graph and by the cluster configuration. Deviations from the expected behavior should raise a warning on possible hardware or software failures.  We performed experiments that show how \sys effectively conveys the heartbeats of some algorithms. This feature can be used both for a visual confirmation of a successful execution and for didactic purposes.
\begin{table}[t]\label{tab:profiling}
\centering
\caption{Resource profiling for \texttt{SSSP} and \texttt{GI} on graph \texttt{cti}.}
\begin{tabular}{|>{\centering\arraybackslash}m{0.07\columnwidth}|>{\centering\arraybackslash}m{0.85\columnwidth}|}
\hline
 \texttt{SSSP} & 
\setlength\fboxrule{0pt}\fbox{\includegraphics[width=0.85\columnwidth]{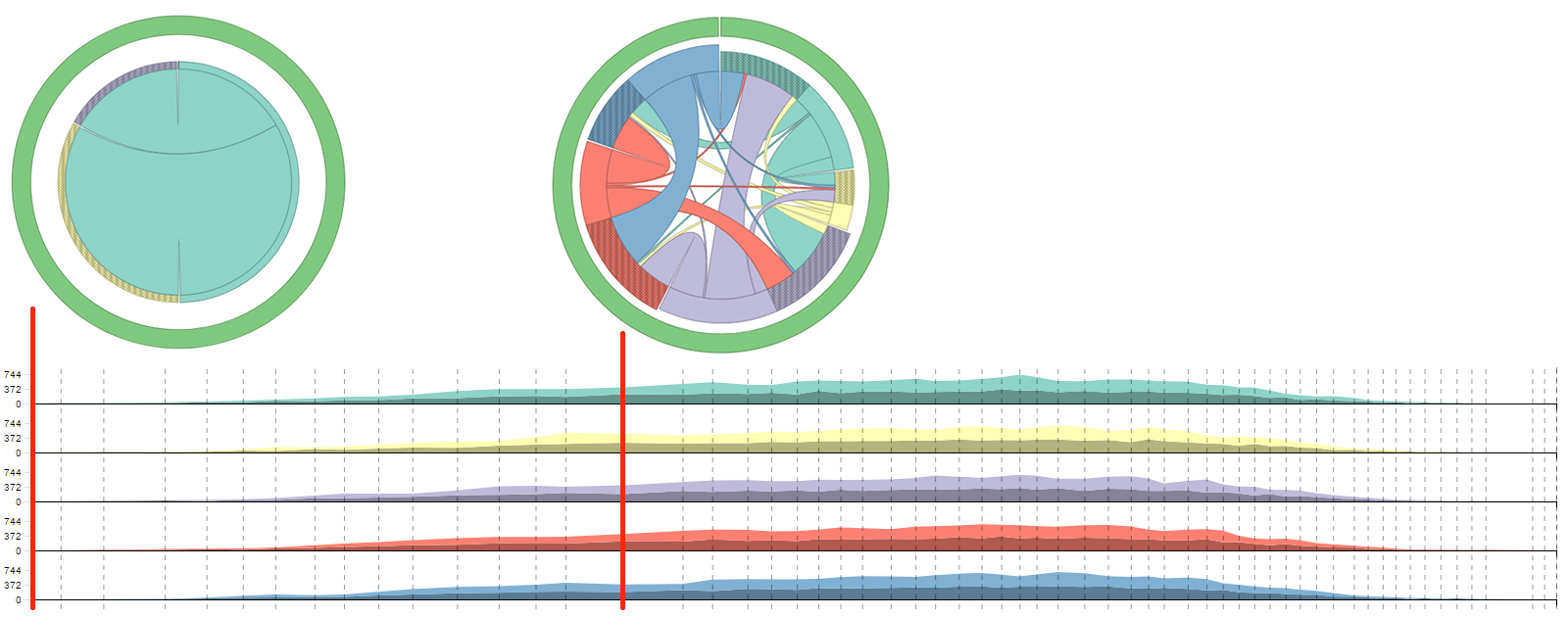}}
  \\
 \hline
 \texttt{GI} &  
\setlength\fboxrule{0pt}\fbox{\includegraphics[width=0.85\columnwidth]{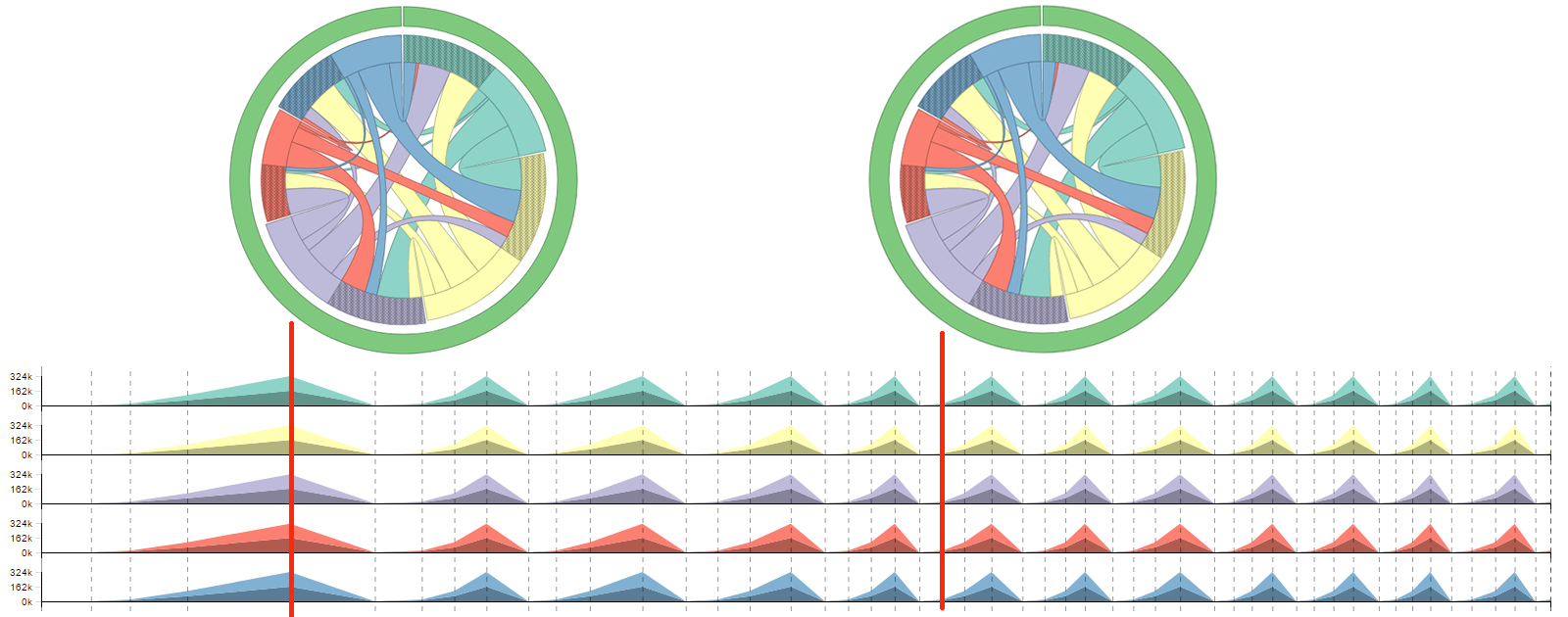}} 
 \\
 \hline
\end{tabular}
\end{table}
We considered four algorithms: \texttt{Single-Source Shortest-Path} (\texttt{SSSP})  and \texttt{Page Rank} (\texttt{PR})~\cite{DBLP:conf/sigmod/MalewiczABDHLC10} are well-known graph algorithms, available in the set of examples provided by the Apache Giraph library; \texttt{GILA} (\texttt{GI})~\cite{DBLP:conf/gd/ArleoDLM16} and \texttt{MultiGILA} (\texttt{MGI})~\cite{DBLP:journals/isci/ArleoDLM17} are TLAV implementations of a force-directed algorithm and of a multilevel force-directed algorithm, respectively. We ran these algorithms on two graphs, \texttt{cti} and \texttt{Gnutella31}. The first is a mesh with $16,840$ vertices and $48,232$ edges, while the second is a peer-to-peer communication network with $62,686$ vertices and $147,892$ edges. Table~\ref{tab:profiling} refers to graph \texttt{cti}. It shows the small multiples representing the exchanged messages (with a hierarchy aggregation at the host level, and after filtering out some hosts with lower traffic), and two representative snapshots of the chord diagram 
The traffic load of \texttt{SSSP} follows a Gaussian-like trend, since the algorithm is based on a flooding technique that reaches its peak when all vertices know their shortest distance from the source vertex. From the first chord diagram, one can see that there is only one host that generates messages in the first superstep, which means that this host contains the source vertex. The messages of \texttt{GI} follows a periodic pattern, where each period represents a controlled flooding  in which the coordinates of a vertex $u$ are broadcast to all vertices within a fixed topological distance from $u$. The chord diagrams at different supersteps look very similar, which tells that the percentages of traffic exchanged between pairs of hosts are stable, even if the total number of messages changes. The traffic load of \texttt{PR} is flat, as the algorithm is based on a set of identical supersteps in which each vertex updates its internal status and communicates with all its neighbors. The chord diagram does not change among different supersteps, as a further witness of this constant behavior. Algorithm \texttt{MGI} alternates computation phases with a periodic trend and phases with flat trend, as a consequence of the multilevel scheme. For example the initial supersteps (concerned with the coarsening phase of algorithm) are very short and generate few messages; the corresponding chord diagrams highlight unbalanced links, due to the fact that only some vertices of the graph are activated in this phase of the algorithm. 

\begin{figure}[t]
	\centering
	\subfigure[]{\includegraphics[width=0.48\columnwidth]{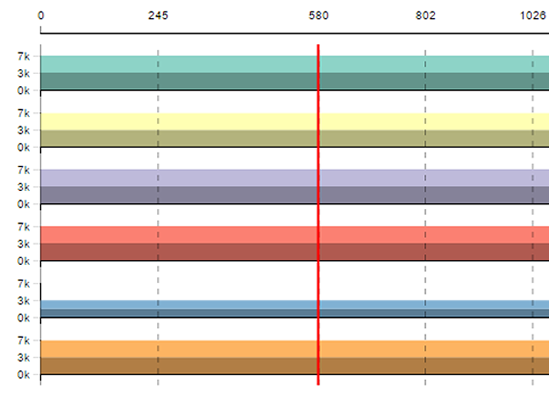}}
	\hfil
	\subfigure[]{\includegraphics[width=0.48\columnwidth]{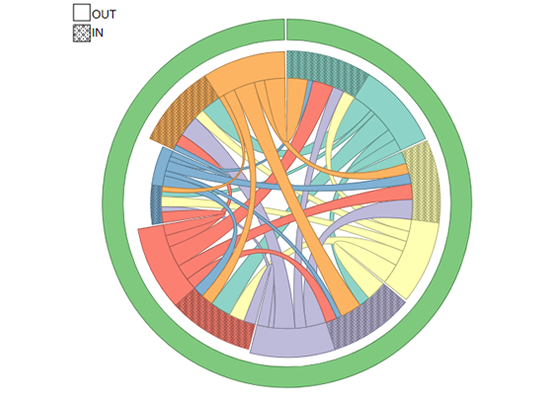}}
	\caption{{\bf Scenario 2:} Anomalous message pattern for algorithm \texttt{SSSP}. Detail of the (a) Trend View and of the (b) Frame View with hierarchy aggregation at the host level.}\label{fig:exp-amp}
\end{figure}

\smallskip\noindent{\bf Scenario 2: Anomalous message patterns.}
%
%
A deviation from the expected heartbeat of an algorithm  should warn the user of a possible issue in the computation. To see this, we injected a bug in the \texttt{SSSP} algorithm and we ran a new experiment. According to the algorithm, if during a superstep there is a vertex $u$ that decreases its best-known distance from the source vertex, then $u$ sends a message to all its neighbors. We added a piece of code that delivers messages to the neighbors of $u$ also if its best-known distance does not change. This causes unnecessary messages, but does not affect the correctness of the algorithm; thus, such a bug would not be discovered by just looking at the output of the computation.  
From the Trend View, the user immediately observes a flat trend of messages, which deviates from the expected Gaussian-like heartbeat 
(see \autoref{fig:exp-amp}).
The chord diagram shows that there are no overloaded links, i.e., the anomalous messages are distributed among the hosts. This confirms that the problem comes from an implementation bug, rather than from a hardware issue.

\smallskip\noindent{\bf Scenario 3: Slow computing units.} Due to the synchronization barriers between supersteps, if a computing unit is significantly slower than the others, it causes a bottleneck for the entire computation. Since the resource management is transparent to the user, such an event is difficult to spot by using default tools such as the Hadoop dashboard and the Giraph counters. In contrast, a slow computing unit can be easily detected in our Trend View. Also, since the problem is usually due to a faulty or overloaded host, an aggregation at the host level may expose the problem.

\begin{figure}[t]
 	\centering
 	\subfigure[]{\includegraphics[width=0.48\columnwidth]{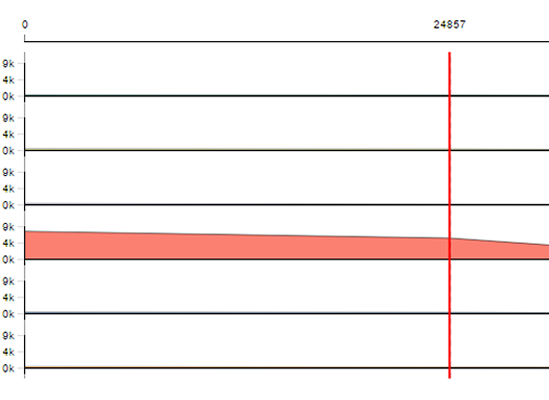}\label{fig:exp-scu-1}}
 	\hfil
 	\subfigure[]{\includegraphics[width=0.48\columnwidth]{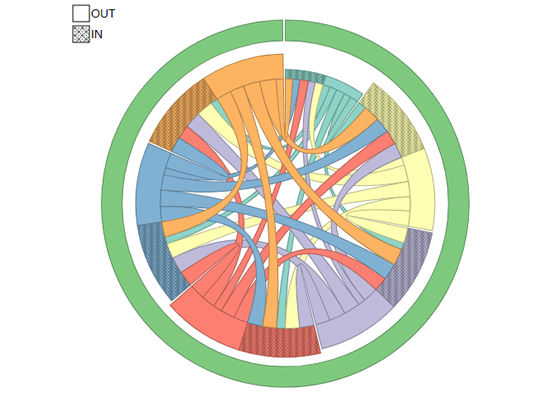}\label{fig:exp-scu-2}}
 	\caption{{\bf Scenario 3:} Slow host for algorithm \texttt{PR}. Detail of the (a) Trend View and of the (b) Frame View with hierarchy aggregation at the host level. }\label{fig:exp-scu}
 \end{figure}

We ran the \texttt{PR} algorithm on the \texttt{4elt} graph (a mesh with $15,607$ vertices and $45,878$ edges). We used our local cluster, whose hosts run within a virtualized environment. We limited the percentage of usable CPU for one of them (while keeping the virtualized hardware the same for all the hosts). The Trend View 
clearly shows the existence of a host whose running time is way higher than the others (indeed, the others are barely visible). Also, the Frame View 
 shows that the slow host (red) handles an amount of messages similar to that of the others. Hence, the poor performance cannot be accounted to a difference in the traffic load, but should be searched in the host conditions (see \autoref{fig:exp-scu}).
 
\section{Discussion and Future Work}\label{se:discussion}
We presented \sys, the first visual profiler for distributed algorithms on Pregel-like graph processing systems, and showed that it can be used in several situations to detect different computation- and infrastructure-related issues.
%
One limitation of \sys is concerned with the Frame View, that requires the usage of filters and/or aggregations if more than a few tens of vertices need to be displayed. This is due to fact that the chord diagram suffers from edge clutter. Although it is uncommon to have more than a few tens of computers allocated for a single computation, one can think of 
investigating alternative graph visualizations, such as matrix-based ones (see, e.g.,~\cite{DBLP:journals/tvcg/BehrischBHDRFS17,zame2008,DBLP:journals/tvcg/HenryFM07}), to improve scalability in our application domain. We also plan to extend \sys with the possibility of executing temporal queries~\cite{Hochheiser:2004}, and of aggregating sequences of supersteps by computation phase. In addition, it would be interesting to collect events from the cluster's resource manager, to detect possible failures of the resource containers. 



{\small \bibliography{paper}}
\bibliographystyle{splncs03}

\clearpage

\appendix

\section*{Appendix}

\begin{figure}[h!]
	\centering
	\includegraphics[width=1.3\columnwidth,angle=90]{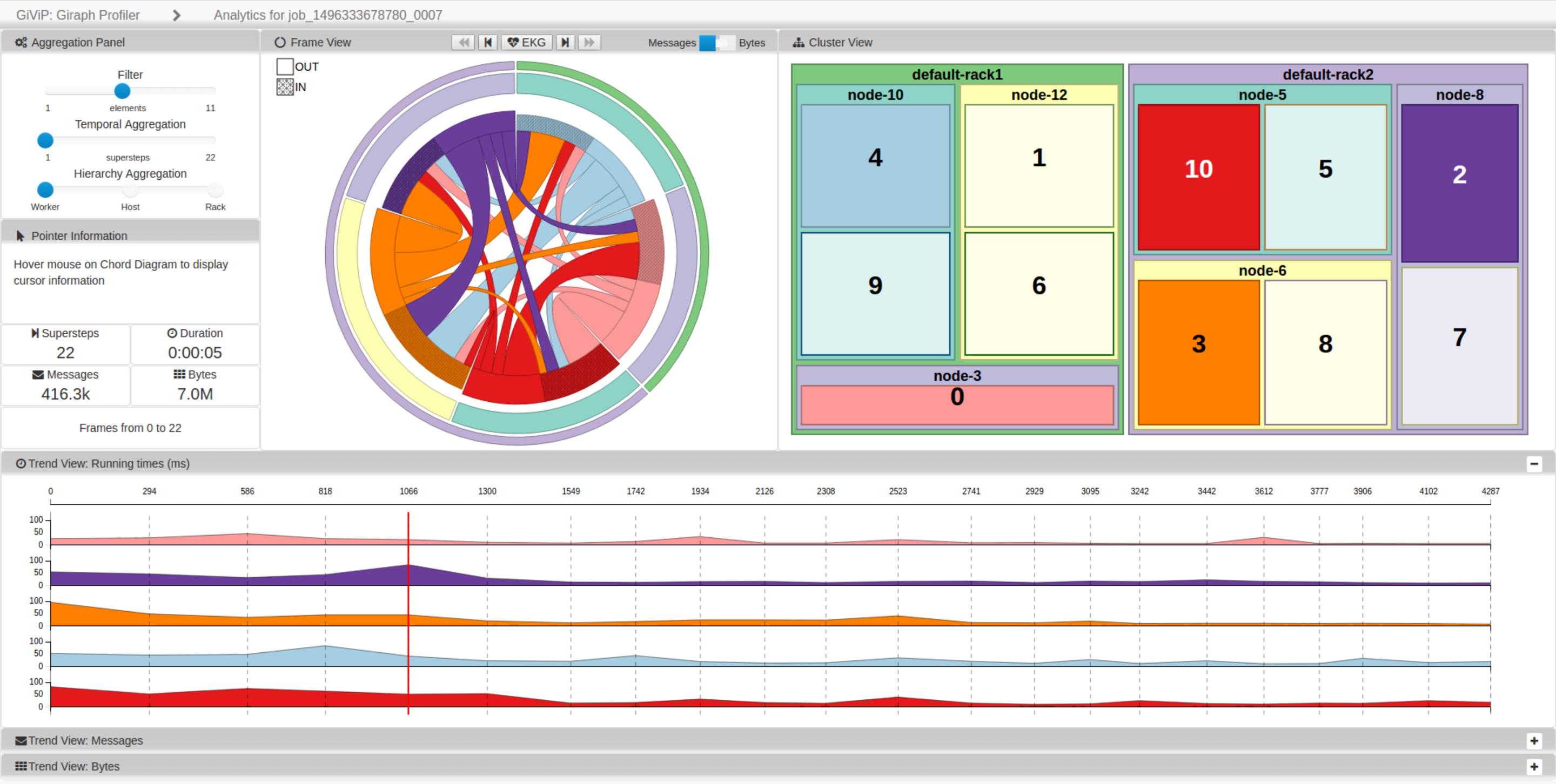}
	\caption{A larger view of Figure \ref{fig:givip}}
	\label{fig:label}
\end{figure}

\section{Additional Material for Section~\ref{sse:visualization}}\label{ap:visualization}

\smallskip\noindent{\em Edge crossing minization heuristics.} As already discussed, we use a variant of the heuristic by Baur and Brandes~\cite{DBLP:conf/wg/BaurB04} to minimize edge crossings in the chord diagram of the Frame View. Our algorithm deals with the constraints imposed by the inclusion tree $T$ and with (dynamic) edge weights. We first briefly recall the algorithm in~\cite{DBLP:conf/wg/BaurB04}.

The algorithm in~\cite{DBLP:conf/wg/BaurB04} iteratively constructs a  permutation of the vertices by adding a vertex per time in a greedy order. The permutation is then used to arrange the vertices around the circle (in a circular layout, edge crossings only depend on the circular order of the vertices). At the generic step, the next vertex $v$ to be inserted is chosen as the one having the minimum number of unplaced neighbors in the graph: $v$ is inserted either at the beginning or at the end of the sequence, based on the position that yields the smaller increment of edge crossings. Then, a \emph{sifting} procedure is applied to further lower the number of edge crossings. 

Our algorithm takes as input the graph $G_{1k}$, where $k$ is the number of supersteps of the computation, and computes a unique circular order of the vertices, used for the visualization of all graphs $G_{ij}$. This is crucial for the user mental map preservation, especially when the visualization changes due to filtering and/or aggregations. The main differences with the algorithm in~\cite{DBLP:conf/wg/BaurB04} are as follows. First, each crossing is assigned a weight equal to the sum of the weights associated with the two crossing edges, and we aim at minimizing the total weight of the edge crossings, rather than their number. The weight of an edge $(u,v)$ of $G_{1k}$ is either $m_{1k}(uv)$ or $b_{1k}(uv)$, based on the user choice. Second, as explained before, vertices corresponding to workers of the same host, as well as those corresponding to hosts of the same rack, must form a contiguous subsequence. This can be achieved by only allowing a vertex to be placed at the beginning or at the end of the subsequence corresponding to its host, and by arranging consecutively the subsequences of hosts that are part of the same rack. Also, in the final sifting procedure each vertex can be moved only within the subsequence of its host. 

\clearpage

\section{Additional Material for Section~\ref{se:evaluation}}\label{ap:evaluation}

\begin{table}[h]\label{tab:profiling-cti-2}
	\centering
	\caption{Resource profiling for \texttt{PR} and \texttt{MGI} on graph \texttt{cti}.}
	\begin{tabular}{|>{\centering\arraybackslash}m{0.07\columnwidth}|>{\centering\arraybackslash}m{0.9\columnwidth}|}
		\hline
		\texttt{PR} & 
		\setlength\fboxrule{0pt}\fbox{\includegraphics[width=0.9\columnwidth]{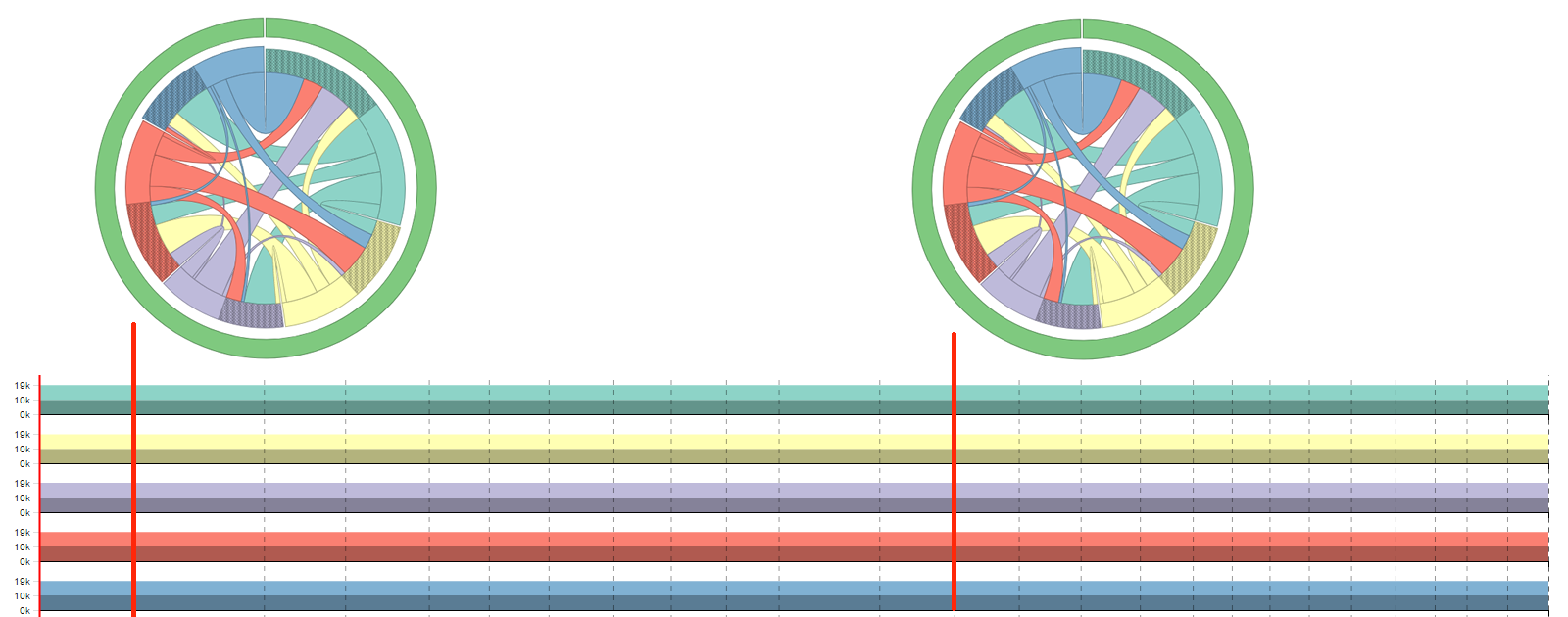}} 
		\\
		\hline
		\texttt{MGI} &   
		\setlength\fboxrule{0pt}\fbox{\includegraphics[width=0.9\columnwidth]{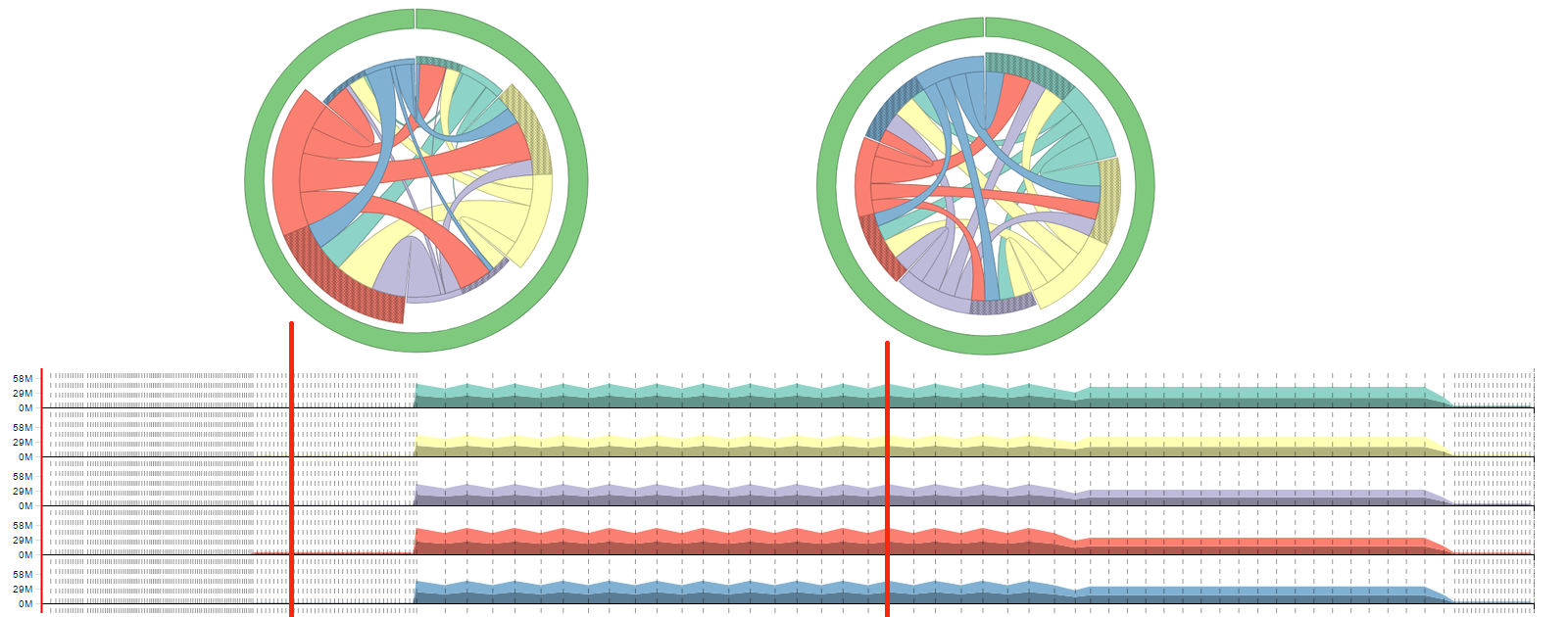}} 
		\\
		\hline
	\end{tabular}
\end{table}

\begin{table}[t]\label{tab:profiling-2}
	\centering
	\caption{Resource profiling for graph \texttt{Gnutella31}.}
	\begin{tabular}{|>{\centering\arraybackslash}m{0.07\columnwidth}|>{\centering\arraybackslash}m{0.9\columnwidth}|}
		\hline
		\texttt{SSSP} & 
		\setlength\fboxrule{0pt}\fbox{\includegraphics[width=0.89\columnwidth]{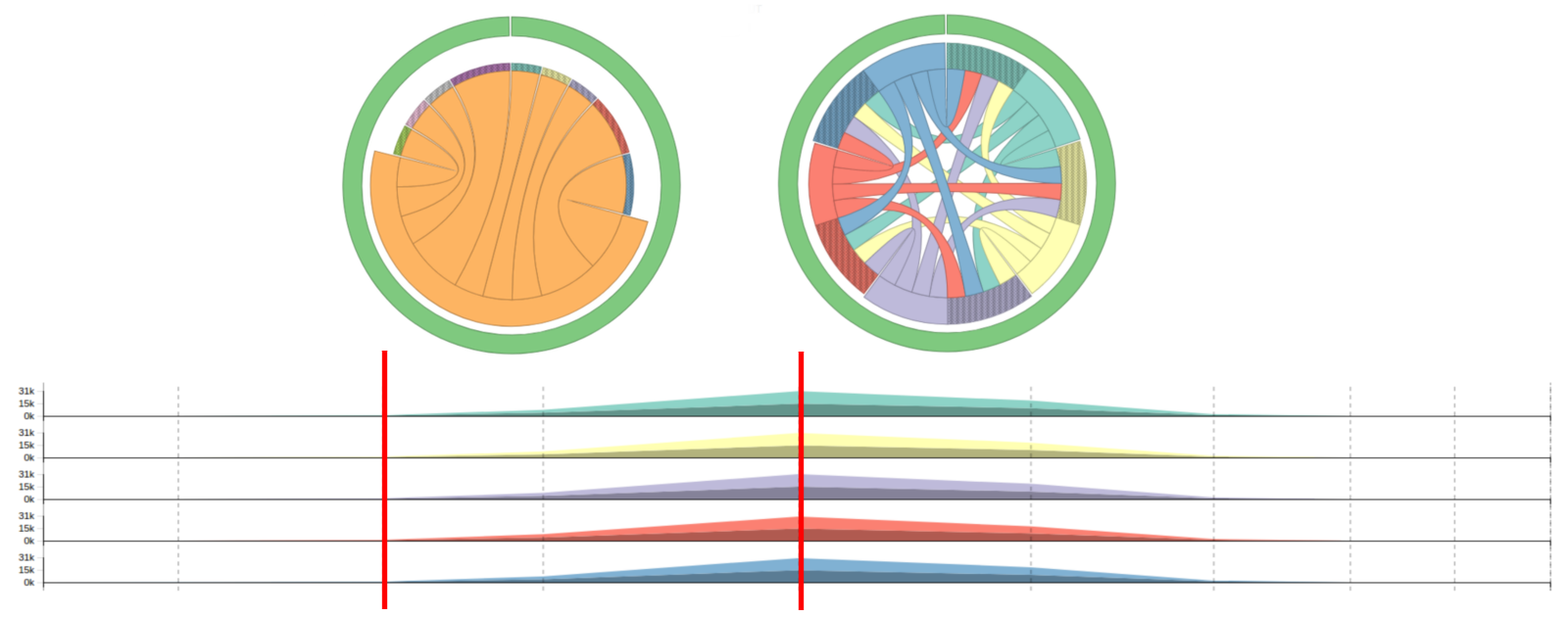}}
		\\
		\hline
		\texttt{PR} & 
		\setlength\fboxrule{0pt}\fbox{\includegraphics[width=0.89\columnwidth]{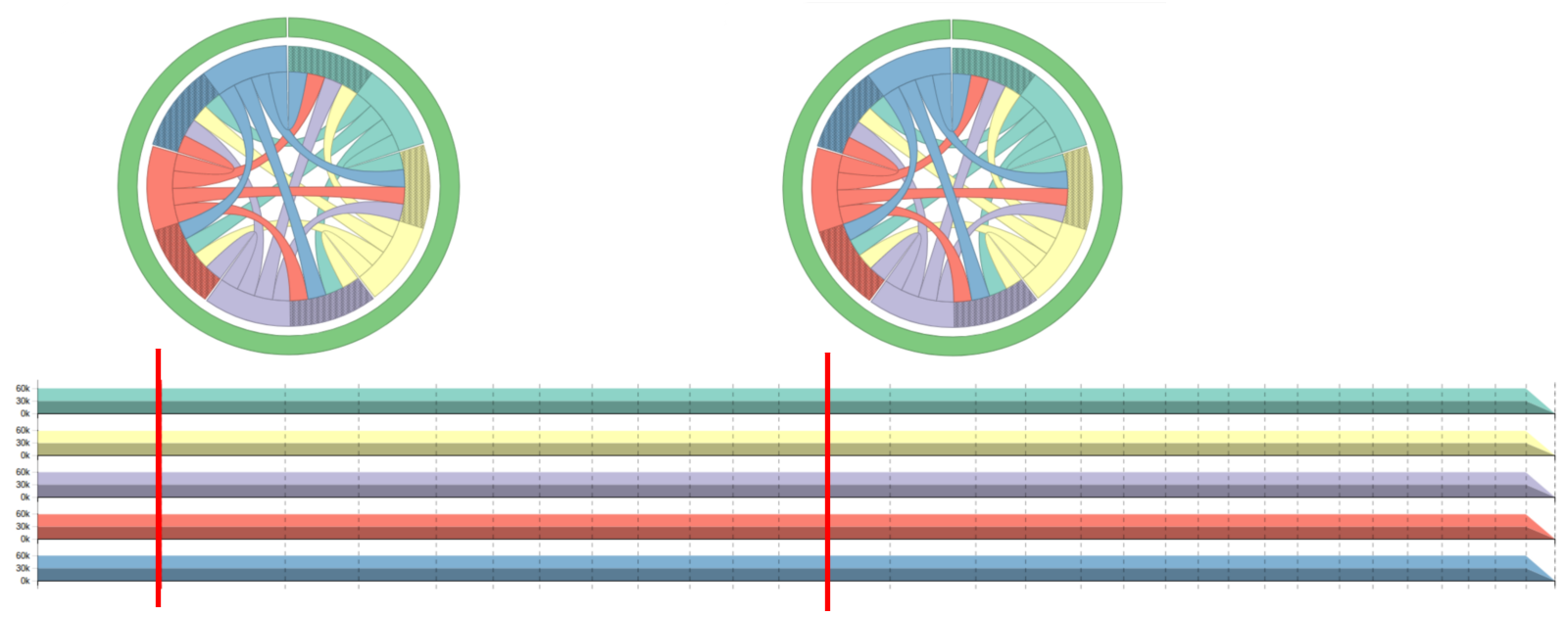}} 
		\\
		\hline
		\texttt{GI} &  
		\setlength\fboxrule{0pt}\fbox{\includegraphics[width=0.89\columnwidth]{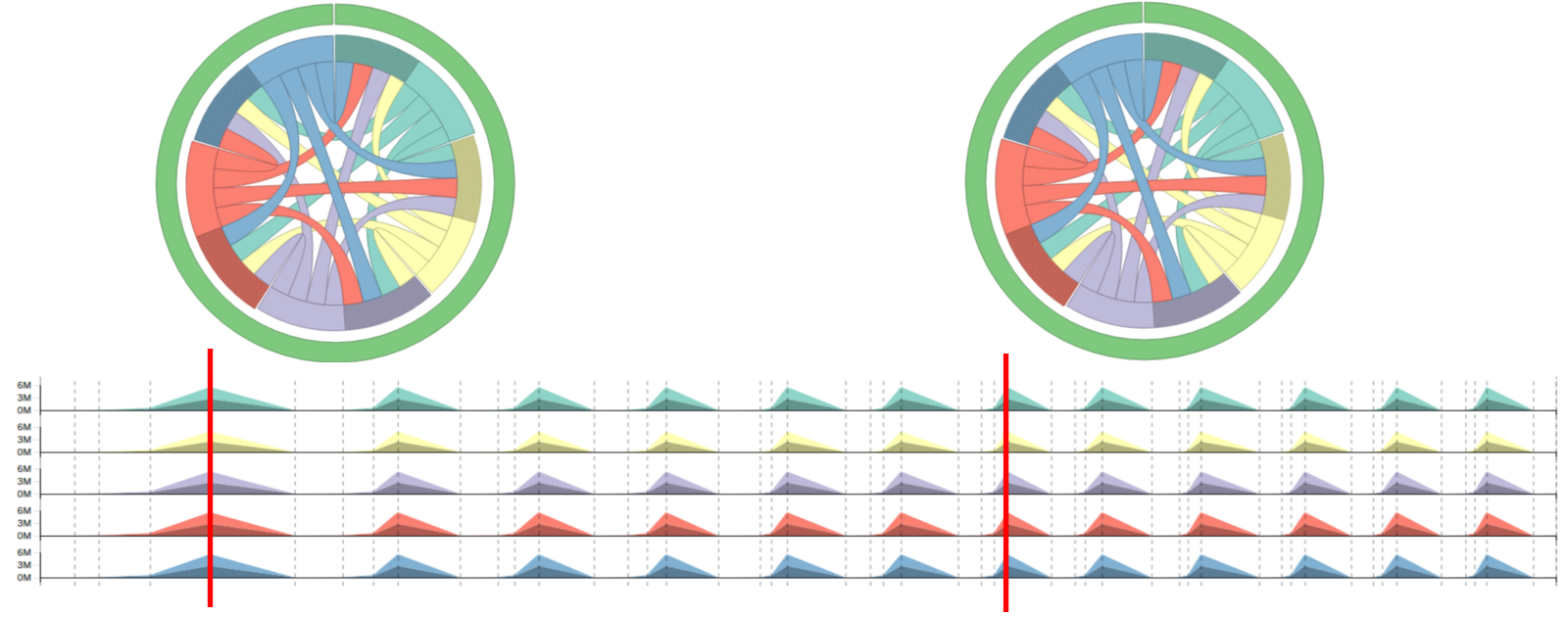}} 
		\\
		\hline
		\texttt{MGI} &   
		\setlength\fboxrule{0pt}\fbox{\includegraphics[width=0.89\columnwidth]{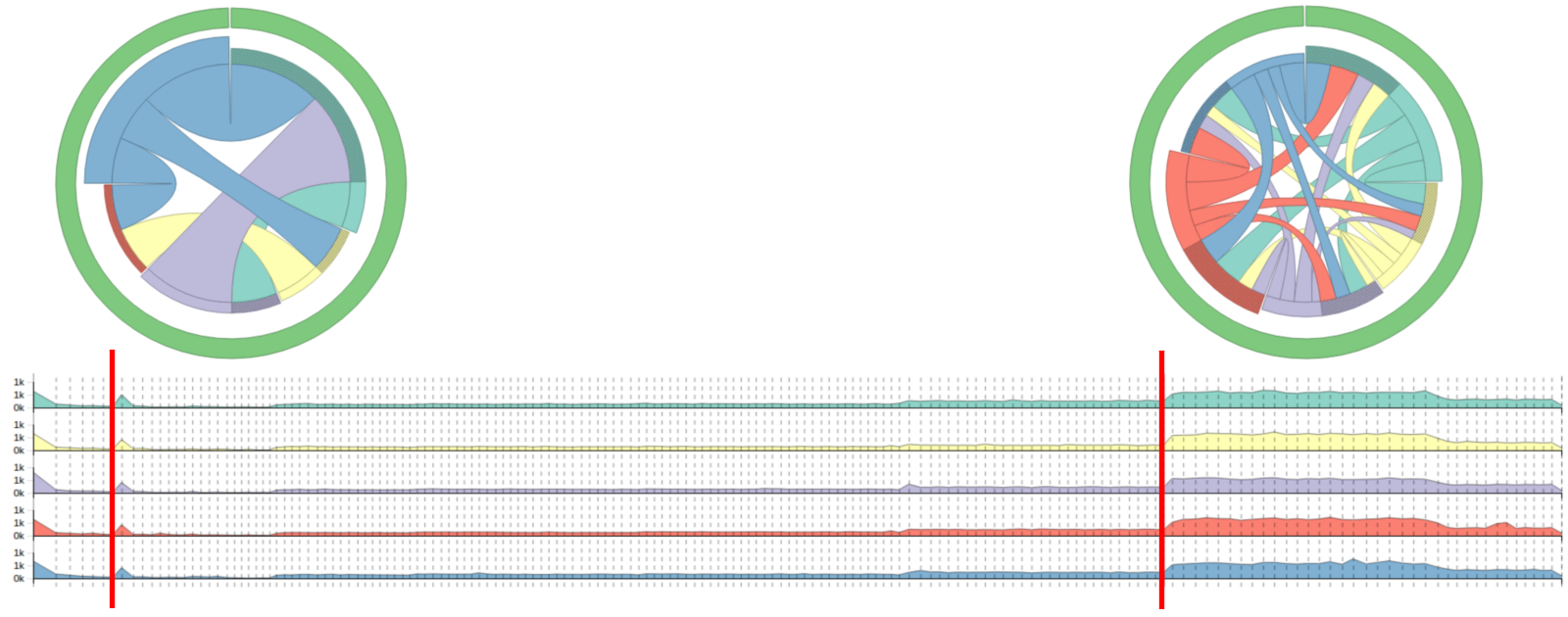}} 
		\\
		\hline
	\end{tabular}
\end{table}
\clearpage

\smallskip\noindent{\bf Scenario 4: Unbalanced partitions.}

While developing the \texttt{GI} and \texttt{MGI} algorithms~\cite{DBLP:conf/gd/ArleoDLM16,DBLP:journals/isci/ArleoDLM17}, we were able to dramatically reduce the running times of these algorithms by replacing the default hash-based partitioning strategy of Giraph with a more sophisticated one, called \texttt{Spinner}~\cite{DBLP:conf/icdcs/VaqueroCLM14}. The algorithm in~\cite{DBLP:conf/icdcs/VaqueroCLM14} computes partitions with increased locality, i.e., such that the number of edges in the input graph having their end-vertices in different workers is minimized. As a consequence, the amount of messages exchanged between different units is reduced, with a clear benefit for the overall performance of our algorithms. 

\begin{figure}[h]
	\centering
	\subfigure[]{\includegraphics[width=0.48\columnwidth]{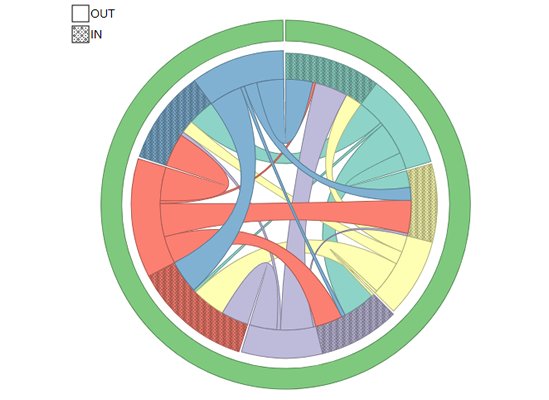}\label{fig:exp-up-a}}
	\hfil
	\subfigure[]{\includegraphics[width=0.48\columnwidth]{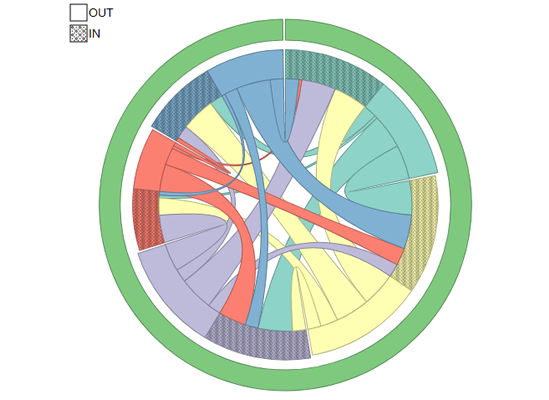}\label{fig:exp-up-b}}
	\caption{{\bf Scenario 4:} The Trend View showing one frame of a computation of algorithm \texttt{GI} using (a) the default hash-based partitioning and (b) the \texttt{Spinner} partitioning.}\label{fig:exp-up}
\end{figure}

We ran algorithm \texttt{GI} on graph \texttt{Gnutella31} with the hash-based partitioning and with the \texttt{Spinner} partitioning. In the first case, the chord diagram shows that there is a heterogeneity in the amount of messages exchanged by different pairs of hosts
(Fig.~\ref{fig:exp-up-a}).
In the second case, the chord diagram shows smaller ribbons (i.e., fewer exchanged messages) and thicker circle arcs (i.e., increased internal traffic), as a witness of the improved locality (Fig.~\ref{fig:exp-up-b}).
Figure~\ref{fig:exp-up-a} shows the chord diagram for algorithm \texttt{GI} on graph \texttt{Gnutella31} and hierarchy aggregation at the host level, when using the hash-based partitioning. It is possible to observe that there is a heterogeneity in the amount of messages exchanged by different pairs of hosts. In contrast, Fig.~\ref{fig:exp-up-b} shows the same algorithm with the same graph but with the \texttt{Spinner} partitioning. The chord diagram shows smaller ribbons (i.e., fewer exchanged messages) and thicker circle arcs (i.e., increased internal traffic), as a witness of the improved locality.


\end{document}